\documentstyle[aps,preprint]{revtex}
\def\beq{\begin{eqnarray}}
\def\eed{\end{eqnarray}}

\begin{document}
\draft

\title{Optical Conductivity in the Copper Oxide Materials}
\author{Zhongbing Huang}
\address{Department of Physics, Beijing Normal University, Beijing
 100875, China and \\
Department of Physics, The Chinese University of Hong Kong, Hong Kong, China}
\author{Shiping Feng\footnote{Regular Associate Member of the Abdus Salam 
International Centre for Theoretical Physics}}
\address{International Centre for Theoretical Physics, P. O. Box 586,
34100 Trieste, Italy and \\
$^{\#}$Department of Physics, Beijing Normal University, Beijing
 100875, China}
\maketitle
\begin{abstract}
The frequency- and temperature-dependent optical conductivity of the
copper oxide materials in the underdoped and optimal doped regimes are
studied within the $t$-$J$ model. The conductivity spectrum shows the
unusual behavior at low energies and anomalous midinfrared peak in the
low temperatures. However, this midinfrared peak is severely depressed
with increasing temperatures, and vanishes at higher temperatures.
\end{abstract}
\narrowtext

After ten years of intense experimental study of the copper oxide
superconductors, a significant body of reliable and reproducible data
has been accumulated by using many probes \cite{n1,n2}, which shows
that the most remarkable expression of the nonconventional physics of
copper oxide materials is found in the normal-state \cite{n1,n2}. The
normal-state properties exhibit a number of anomalous properties in
the sense that they do not fit in the conventional Fermi-liquid theory,
and some properties mainly depend on the extent of dopings \cite{n1,n2}.
Among the striking features of the anomalous properties stands out the
extraordinary optical conductivity \cite{n3}. The frequency- and
temperature-dependent optical conductivity is a powerful probe for
systems of interacting electrons, and provides very detailed
information of the excitations, which interact with carriers in the
normal-state and might play an important role in the superconductivity
\cite{n3}. The optical conductivity of the copper oxide materials in the
underdoped and optimally doped regimes has been extensively studied
\cite{n3,n4,n5,n61}, and the experimental results indicate that the
optical conductivity spectrum shows unusual behavior at low energies and
anomalous midinfrared band in the charge-transfer gap, which is
inconsistent with the conventional electron-phonon scattering mechanism.

Since the undoped copper oxide materials are antiferromagnetic Mott
insulators, and upon doping with holes in the copper oxide sheets, the
antiferromagnetic long-range-order (AFLRO) disappears and
superconductivity emerges as the ground state \cite{n2}, many
researchers believe that the essential physics is contained in the
doped antiferromagnets \cite{n6,n7}, which may be effectively described
by the two-dimensional (2D) $t$-$J$ model acting on the space with no
doubly occupied sites. In spite of its simple form the $t$-$J$ model
proved to be very difficult to analyze, analytically as well as
numerically, because of the electron single occupancy on-site local
constraint. The local nature of the constraint is of primary importance,
and its violation may lead to some unphysical results \cite{n9}.
Recently a fermion-spin theory based on the charge-spin separation
has been proposed to incorporate this constraint \cite{n10}. The main
advantage of this approach is that the electron on-site local
constraints can be treated exactly in analytical calculations. Within
the fermion-spin theory, we \cite{n11} have shown that AFLRO vanishes
around doping $\delta=5\%$ for an reasonable value of the parameter
$t/J=5$. The mean-field theory in the underdoped and optimally doped
regimes without AFLRO has been developed \cite{n12}, which has been
applied to study the
photoemission, electron dispersion and electron density of states in the
copper oxide materials, and the results are qualitatively consistent with
experiments and numerical simulations. In this paper, we consider
fluctuations around this mean-field solution to study the optical
conductivity, and show that the result within the fermion-spin
formalism exhibits a  behavior similar to that seen in the experiments
and numerical simulations.

We begin with the $t$-$J$ model defined on a square lattice,
\begin{equation}
H = -t\sum_{\langle ij\rangle\sigma}C^{\dagger}_{i\sigma}C_{j\sigma} + h.c.
- \mu \sum_{i\sigma}C^{\dagger}_{i\sigma}C_{i\sigma} +
J\sum_{\langle ij\rangle}{\bf S}_{i}\cdot {\bf S}_{j} ,
\end{equation}
where $C^{\dagger}_{i\sigma}$ ($C_{i\sigma}$) are the electron creation
(annihilation) operators,
${\bf S}_{i}=C^{\dagger}_{i}{\bf \sigma} C_{i}/ 2$ are spin operators
with ${\bf \sigma}=(\sigma_{x},\sigma_{y},\sigma_{z})$ as Pauli matrices,
$\mu$ is the chemical potential, and the summation $\langle ij\rangle$
is carried over nearest nonrepeated bonds. Equation (1) is subject
to an important constraint that a given site can not be occupied by
more than one electron
$\sum_{\sigma}C^{\dagger}_{i\sigma}C_{i\sigma}\leq 1$. This local
constraint is satisfied even in the mean-field approximation
within the fermion-spin transformation \cite{n10,n12},
\begin{eqnarray}
C_{i\uparrow}=h^{\dagger}_{i}S^{-}_{i},~~~~
C_{i\downarrow}=h^{\dagger}_{i}S^{+}_{i},
\end{eqnarray}
where the spinless fermion operator $h_{i}$ keeps track of the charge
(holon), while the pseudospin operator $S_{i}$ keeps track of the spin
(spinon). In this case, the $t$-$J$ model (1) can be rewritten in the
fermion-spin representation as,
\begin{eqnarray}
H = -t\sum_{\langle ij\rangle}h_{i}h^{\dagger}_{j}(S^{+}_{i}S^{-}_{j}
+S^{-}_{i}S^{+}_{j}) +  h. c.
- \mu \sum_{i}h^{\dagger}_{i}h_{i} +
J\sum_{\langle ij\rangle}(h_{i}h^{\dagger}_{i})({\bf S}_{i}
\cdot {\bf S}_{j})(h_{j}h^{\dagger}_{j}) ,~~~~~
\end{eqnarray}
with $S^{+}_{i}$ and $S^{-}_{i}$ as pseudospin raising and lowering
operators, respectively. It is obvious that there is an interaction
between spinons and holons in the Hamiltonian (3). The spinon and
holon may be separated at the mean-field level, but they are strongly
coupled beyond MFA due to fluctuations.

The mean-field theory within the fermion-spin formalism in the underdoped
and optimally doped regimes without AFLRO has been developed
\cite{n12}, and the mean-field spinon Green's function
$D^{(0)}(i-j,\tau-\tau')=-\langle T_\tau S^{+}_{i}(\tau)S^{-}_{j}(\tau')
\rangle_0$ and mean-field holon Green's function
$g^{(0)}(i-j,\tau-\tau')=-\langle T_\tau h_{i}(\tau)h^{\dagger}_{j}(\tau')
\rangle_0$ have been evaluated \cite{n12} as,
\begin{eqnarray}
D^{(0)}({\bf k},\omega)={\Delta [(2\epsilon\chi_{z}+\chi)\gamma_{k}-
(\epsilon\chi+2\chi_{z})]\over 2\omega (k)}\left (
{1\over \omega -\omega (k)}-{1\over \omega +\omega (k)}\right ),
\end{eqnarray}
\begin{eqnarray}
g^{(0)}({\bf k},\omega)={1\over \omega-\xi_k},
\end{eqnarray}
respectively, where
$\gamma_{{\bf k}}=(1/Z)\sum_{\eta}e^{i{\bf k}\cdot\hat{\eta}}$,
$\hat{\eta }=\pm \hat{x},\pm \hat{y}$,
$\Delta =2ZJ_{eff}$, $J_{eff}=J[(1-\delta)^2-\phi^2]$,
$\epsilon=1+2t\phi/J_{eff}$, $Z$ is the number of the nearest neighbor
sites, the mean-field spinon excitation spectrum
\begin{eqnarray}
\omega^{2}(k)=\Delta^{2}\left (\alpha\epsilon(\chi_{z}\gamma_{k}-
{1\over Z}\chi )(\epsilon\gamma_{k}-1)+[\alpha C_{z}+{1\over 4Z}
(1-\alpha)](1-\epsilon\gamma_{k})\right ) \nonumber \\
+\Delta^{2}\left ({1\over 2}\alpha\epsilon\chi\gamma_{k}(\gamma_{k}
-\epsilon) + {1\over 2}\epsilon[\alpha C+{1\over 2Z}
(1-\alpha)](\epsilon-\gamma_{k}) \right ),
\end{eqnarray}
and the mean-field holon excitation spectrum $\xi_k=2Z\chi t\gamma_k-\mu$,
with the spinon correlation functions $\chi=\langle S_{i}^{+}S_{i+\eta }^{-}
\rangle =\langle S_{i}^{-}S_{i+\eta}^{+}\rangle$,
$\chi_{z}=\langle S_{i}^{z}S_{i+\eta}^{z}\rangle$, $C=(1/Z^{2})
\sum_{\eta ,\eta ^{\prime }}\langle S_{i+\eta }^{+}S_{i+\eta
^{\prime}}^{-}
\rangle$, $C_{z}=(1/Z^{2})\sum_{\eta ,\eta ^{\prime }}\langle
S_{i+\eta }^{z}S_{i+\eta ^{\prime }}^{z} \rangle$, and
holon particle-hole order parameter
$\phi =\langle h_{i}^{\dagger}h_{i+\eta }\rangle$. In order
not to violate the sum rule of the correlation function
$\langle S^{+}_{i}S^{-}_{i}\rangle=1/2$ in the case without AFLRO, the
important decoupling parameter $\alpha$ has been introduced in the
mean-field calculation \cite{n12}, which can
be regarded as the vertex correlations \cite{n13}. The order parameters
$\chi$, $C$, $\chi_z$, $C_z$, $\phi$ and chemical potential $\mu$ have
been determined \cite{n12} by solving the self-consistent equations.

For discussing the optical conductivity, we now need to consider the
fluctuations around the above mean-field solution. In the fermion-spin
framework, an electron is represented by the product of a holon and a
spinon, then the external field can only be coupled to one of them.
Ioffe and Larkin \cite{n14} have shown that the physical conductivity
$\sigma(\omega)$ is given by,
\begin{eqnarray}
\sigma^{-1}(\omega)=\sigma_{h}^{-1}(\omega)+\sigma_{s}^{-1}(\omega),
\end{eqnarray}
where $\sigma_h(\omega)$ and $\sigma_s(\omega)$ are the contributions
to the conductivity from holons and spinons, respectively, and can be
expressed as \cite{n15},
\begin{eqnarray}
\sigma_h(\omega)=-{\rm Im}\Pi_{h}(\omega)/\omega,~~~~
\sigma_s(\omega)=-{\rm Im}\Pi_{s}(\omega)/\omega,
\end{eqnarray}
with $\Pi_h(\omega)$ and $\Pi_s(\omega)$ as holon and spinon
current-current correlation functions,
respectively, and defined as,
\begin{eqnarray}
\Pi_s(\tau-\tau')=-\langle T_\tau j_s(\tau)j_s(\tau')\rangle,~~
\Pi_h(\tau-\tau')=-\langle T_\tau j_h(\tau)j_h(\tau')\rangle,
\end{eqnarray}
where the current densities of spinons and holons are expressed in the
present theoretical framework as,
\begin{eqnarray}
j_{s}=te\phi\sum_{i\eta}\hat\eta (S_{i}^{+}S_{i+\eta}^{-}+
S_{i}^{-}S_{i+\eta}^{+}),\\
j_{h}=2te\chi\sum_{i\eta}\hat\eta h_{i+\eta}^{+}h_{i},
\end{eqnarray}
respectively. In a formal calculation \cite{n161} for the spinon
current-current correlation function we find $\Pi_s=0$. However,
the strongly correlation between holons and spinons is still considered
through the spinon's order parameters $\chi$,
$\chi_z$, $C$ and $C_z$ entering the holon current-current correlation
function, which means that the holon moves in the background of spinons,
and the cloud of distorted spinon background is to follow holons.
Therefore the dressing of the holon by spinon excitations is the key
ingredient in the explanation of the optical conductivity of the copper
oxide materials.

The holon current-current correlation function defined in Eq. (9)
can be rewritten as,
\begin{eqnarray}
\Pi_{h}(i\omega_{n})=-(2te\chi Z)^{2}\frac{1}{N}\sum_{k}\gamma_{sk}^{2}
\frac{1}{\beta}\sum_{i\omega_{m}'}g(k, i\omega_{m}'+i\omega_{n})
g(k, i\omega_{m}'),
\end{eqnarray}
where $i\omega_{n}$ is the Matsubara frequency,
$\gamma_{sk}=(1/2)(\sin k_{x}+\sin k_{y})$, and $g(k, i\omega_{n})$
is the full holon Green's function. In this paper, we consider the
second-order
correction for the holon. The second-order holon self-energy diagram from
the spinon pair bubble has been discussed in Ref. \cite{n161}, and
the result was obtained as,
\begin{eqnarray}
\Sigma_{h}^{(2)}(k,i\omega_{n})=(Zt)^{2}{1\over N^2}\sum_{pp'}
(\gamma_{p'-k}+\gamma_{p'+p+k})^{2}B_{p'}B_{p+p'}\times \nonumber \\
\left ( 2{n_{F}(\xi_{p+k})
[n_{B}(\omega_{p'})-n_{B}(\omega_{p+p'})]-n_{B}(\omega_{p+p'})
n_{B}(-\omega_{p'})\over i\omega_{n}+\omega_{p+p'}-\omega_{p'}-
\xi_{p+k}} \right. \nonumber \\
+{n_{F}(\xi_{p+k})[n_{B}(\omega_{p+p'})-n_{B}(-\omega_{p'})]+
n_{B}(\omega_{p'})n_{B}(\omega_{p+p'})\over i\omega_{n}+\omega_{p'}+
\omega_{p+p'}-\xi_{p+k}} \nonumber \\
\left. -{n_{F}(\xi_{p+k)}[n_{B}(\omega_{p+p'})
-n_{B}(-\omega_{p'})]-n_{B}(-\omega_{p'})n_{B}(-\omega_{p+p'}) \over
i\omega_{n}-\omega_{p+p'}-\omega_{p'}-\xi_{p+k}}\right ),
\end{eqnarray}
where $B_k=ZJ_{eff}[(2\epsilon \chi_z+\chi)
\gamma_{k}-(\epsilon \chi +2\chi_z)]/\omega_{k}$,
$n_{F}(\xi_{k})$ and $n_{B}(\omega_{k})$ are the Fermi and Bose
distribution functions, respectively.
Then the full holon Green's function is obtained,
\begin{eqnarray}
g(k,i\omega_{n})={1\over g^{(0)-1}(k,i\omega_{n})-\Sigma_{h}^{(2)}
(k,i\omega_{n})}={1\over i\omega_{n}-\xi_{k}-\Sigma_{h}^{(2)}
(k,i\omega_{n})}.
\end{eqnarray}
The above full holon Green's function $g(k,i\omega_{n})$ can also be
expressed as frequency integrals in terms of the spectral representation,
\begin{eqnarray}
g(k,i\omega_{n})=\int_{-\infty}^{\infty}{d\omega\over 2\pi}
{A_{h}(k,\omega)\over i\omega_{n}-\omega},
\end{eqnarray}
with the holon spectral function $A_{h}(k,\omega)=-2{\rm Im}g(k,\omega)$.
Substituting Eq. (15) into Eq. (12), and evaluating the frequency
summations,  we obtain the optical conductivity from Eqs. (7) and
(8) as,
\begin{eqnarray}
\sigma (\omega )={1\over 2}(2te\chi Z)^2{1\over N}\sum_k\gamma_{sk}^{2}
\int^{\infty}_{-\infty}{d\omega'\over 2\pi}A_{h}(k,\omega'+\omega)
A_{h}(k,\omega'){n_{F}(\omega'+\omega)-n_{F}(\omega') \over \omega}.~~~~
\end{eqnarray}

Although the optical properties of the copper oxide materials are very
complicated, some qualitative features, such as (1) a sharp peak at
$\omega=0$, (2) considerable weight appears inside the charge-transfer
gap of the undoped materials, defining  the midinfrared band, and (3) the
conductivity decays as $\rightarrow 1/\omega$ at low energies, seem to
be common to all copper oxide materials \cite{n3,n17}. In the following,
we study the frequency- and temperature-dependent optical conductivity of
the copper oxide materials in the underdoped and optimally doped regimes.
We have performed a numerical calculation for the optical conductivity
(16) at finite temperatures, and the results with temperature $T=0.2J$
at dopings $\delta=0.06$ (solid line), $\delta=0.10$ (dashed line), and
$\delta=0.15$ (dot line) for the parameter $t/J=2.5$ are plotted in
Fig. 1, where the charge e has been set as the unit. Our low temperature
results show that there is a low-energy peak at $\omega < 0.5t$ separated
by a gap or pseudogap $=0.5t$ from the broad absorption band (midinfrared
band) in the conductivity spectrum. Moreover, The midinfrared spectral
weight is doping dependent, increasing with doping for $0.5t<\omega <2t$
and is nearly independent of doping for $\omega >2t$. In particular, the
midinfrared spectral weight is biased towards the lower energy region with
increased doping. This reflects the increase in the mobile carrier
density,
and indicates that the spectral weight of the midinfrared sideband is
taken from the Drude absorption. Therefore
the spectral weight from both the low-energy peak and midinfrared
sideband represents the actual free-carrier density. These results are in
qualitative agreement with  experiments \cite{n4,n5,n61} and numerical
simulations \cite{n20,n21}.

For further understanding the property of the optical conductivity,
we show $\sigma(\omega)$ at (a) doping $\delta=0.06$
and (b) $\delta=0.15$ for $t/J=2.5$ with temperatures $T=0.2J$ (solid
line), $T=0.4J$ (dashed line), $T=0.6J$ (dot-dashed line), and $T=1.0J$
(dot line) in Fig. 2. In comparison with the low temperature result in
Fig. 1, we find that the conductivity is temperature-dependent
for $\omega <1.5t$ and almost temperature-independent for $\omega >1.5t$.
The peak at $\omega =0$ broadens and decreases in height with increasing
temperatures, and the component in the low-energy region also increases
with increasing temperature. Therefore there is a tendency towards
the Drude-like behavior,
while the midinfrared spectral weight (centered near $\omega\approx 1t$)
is severely suppressed with increasing temperatures, and vanishes at
higher
temperatures ($T>0.5J$), which are also consistent with the numerical
simulations \cite{n22} and experiments \cite{n4,n5,n61,n23}. Although the
midinfrared spectral weight vanishes at higher temperatures, the total
spectral weight of the optical conductivity does not change since the
midinfrared spectral weight has been incorporated into the low-energy
spectral weight, and therefore the sum rule of the optical conductivity
\cite{n24} is still satisfied. In the present fermion-spin theory based on
the charge-spin separation, the basic low-energy excitations are holons and
spinons, but our theoretical results show that the anomalous optical
properties at finite temperature are mainly caused by the charged holons
in the copper oxide sheets, which are strongly renormalized because of
the strong interactions with fluctuations of the surrounding spinon
excitations.

In summary, we have studied the frequency- and temperature-dependent
optical conductivity of the copper oxide materials in the underdoped and
optimally doped regimes within the $t$-$J$ model, and the theoretical
results
of the optical conductivity at finite temperatures are qualitatively
consistent with experiments \cite{n5,n23} and numerical simulations
\cite{n20,n22}. Our optical spectra have been used to extract the dc
conductivity and resistivity \cite{n161}, and the result shows that the
resistivity indeed exhibits a very good linear behavior at low
temperatures in the underdoped and optimally doped regimes.

\acknowledgments
This work was supported by the National Science Foundation Grant No.
19774014, and the State Education Department of China through the
Foundation of the Doctoral Training. The partial support from the
Associateship Scheme of the Abdus Salam International Centre for
Theoretical Physics at Trieste, Italy is also acknowledged.

\begin{figure}
\caption{The optical conductivity at the doping $\delta=0.06$ (solid
line),
$\delta=0.10$ (dashed line), and $\delta=0.15$ (dot line) for the
parameter $t/J=2.5$ with the temperature $T=0.2J$.}
\label{autonum}
\end{figure}

\begin{figure}
\caption{The optical conductivity at (a) the doping $\delta=0.06$ and
(b) $\delta=0.15$ for the parameter $t/J=2.5$ with temperatures $T=0.2J$
(solid line), $T=0.4J$ (dashed line), $T=0.6J$ (dot-dashed line),
$T=1.0J$ (dot line).}
\label{autonum}
\end{figure}

\end{document}